\documentclass[]{article} 
\usepackage[utf8]{inputenc} 
\usepackage{authblk}
\usepackage{hyperref}
\usepackage{bm}
\usepackage{graphicx}
\usepackage{fancybox}
\usepackage{ascmac}
\usepackage{amsmath}
\usepackage{braket}
\usepackage{setspace}
\usepackage{breqn}
\usepackage[a4paper, margin=1in]{geometry} 
\usepackage{cite} 
\usepackage{subcaption}
\usepackage{booktabs}
\usepackage{upgreek} 

\usepackage{ifthen}
\usepackage{ulem} 
\usepackage{xcolor} 
\usepackage{xparse} 

\NewDocumentCommand{\added}{m}{\ifthenelse{\boolean{draft}}{\textcolor{blue}{#1}}{#1}}
\NewDocumentCommand{\deleted}{m}{\ifthenelse{\boolean{draft}}{\textcolor{red}{\sout{#1}}}{}}

\providecommand{\keywords}[1]{\textbf{\textit{Keywords: }} #1}

\title{Time-Dependent Density Functional Theory Simulation for Analyzing Neutralization Process of Hydrogen Ion Injected onto Tungsten Surfaces} 

\author[1]{Yuto Toda}
\author[2,1]{Arimichi Takayama}
\author[2,1]{Atsushi M. Ito}

\affil[1]{Graduate Institute for Advanced Studies, SOKENDAI, Oroshi-cho 322-6, Toki 509-5292, Japan}
\affil[2]{National Institute for Fusion Science, National Institutes of Natural Sciences, Oroshi-cho 322-6, Toki 509-5292, Japan}
\affil[ ]{\texttt{toda.yuto@nifs.ac.jp}}

\date{\today}

\begin{document}

\newboolean{draft}
\setboolean{draft}{false} 

\maketitle
\doublespacing 

\begin{abstract}
We \added{have} performed time-dependent density functional theory simulations for the neutralization process of a hydrogen ion injected at 100eV onto the (110) surface of tungsten material. 
We \added{have} also proposed a method for evaluating the detection probability of electrons in a small region.
This probability is interpreted as \added{that}\deleted{the probability} of detecting hydrogen in each state: positive ion, neutral atom, and negative ion. 
As a result, the probabilities of detecting hydrogen after \added{a}\deleted{the} collision as a positive ion, neutral atom, and negative ion were approximately 30 percent, 50 percent, and 20 percent, respectively.
\end{abstract}


\section{Introduction}

Atomic-scale plasma-wall interaction (PWI) simulations have long used the binary collision approximation (BCA) for physical sputtering\added{and}\deleted{,} molecular dynamics (MD) to incorporate many-body effects such as chemical sputtering, and density functional theory (DFT) to provide more accurate calculations \cite{Wirth2015,Kajita2022,Nordlund2006}. 
Previous PWI simulations have generally employed neutral atoms to substitute for injected ions from\deleted{the}plasma. 
These assumed that the ion from plasma is rapidly neutralized on the surface of materials. 
However, experimental studies \added{have} confirmed that the incident ion is reflected not only as a neutral atom but also as an ion without neutralization or with re-ionization\added{,} following neutralization\cite{Souda1985}. 
\deleted{In addition, a negative ion is produced by collision with incident ions and the surface of the material. }
\deleted{The experimental production ratio of negative hydrogen on the surfaces of various metals has been reported}%
Although there are some theoretical works to explain the neutralization process of an incident ion on a surface \cite{Horiguchi1978,Hagstrum1961,Trubnikov1967,Kitagawa1976}, it is not sufficiently understood.

Pioneering research on ion injection using time-dependent density functional theory (TDDFT)\cite{Runge1984} in plasma-material interaction (PMI) was reported by Miyamoto et al.\cite{Miyamoto2008_1,Miyamoto2008_2}. 
They performed a simulation on the incidence of $\text{Ar}^{7+}$ on graphene and its excited states. 
However, in particular, we simulate the neutralization process of an ion on a surface using TDDFT and analyze the results from a quantum mechanical perspective. 
Suppose we can obtain the probability of detecting exactly one electron around the hydrogen nucleus from the electron wave function after the incident hydrogen collides with the tungsten surface. 
This can be interpreted as the neutralization probability of the incident hydrogen, in this study, we aim to estimate.

\section{Simulation Method}

In order to simulate the neutralization process of ion injected from plasma onto materials, it is necessary to simultaneously solve both the motion of electron wave functions and the motion of nuclei. 
Hence, we introduce a simulation method that combination of TDDFT and MD\cite{Curchod2013}. 
This method applies different treatments of electrons and nuclei. 
The electrons are treated with a wave function and the nuclei classical particles. 
The motion of electron wave functions are calculated using the \added{time-dependent Kohn-Sham} (KS) equation
\begin{align}
i\frac{\partial \psi_j(x,t)}{\partial t} = \hat{H}_{\rm KS} \psi_j(x, t)\label{eq:KS_eq}
\end{align}
based on TDDFT. Here, $\psi_j(x, t)(j = 1, 2, \cdots, n)$ are the $j$-th KS orbital, and $\hat{H}_{\rm KS}$ is the KS Hamiltonian.
Moreover, the motions of the nuclei are simulated using the Newton equation
\begin{align}
    M_{\mathrm{\upalpha}}\frac{d^2 \vec{R}_{\upalpha}}{dt^2}=-\frac{\partial}{\partial \vec{R}_{\mathrm{\upalpha}}}\left\{E_{\rm nn}(\vec R) + E_{\rm KS}[\rho] \right\}\label{eq:Newton_eq}
\end{align}
as classical particles, where $E_{\rm nn}(\vec{R})$ is the interaction energy among all nuclei\cite{Ito2024} and $E_{\rm KS}[\rho]$ is the KS energy as a functional of the electronic density $\rho$.

\begin{figure}[ht]
    \centering
    \includegraphics[width=0.8\textwidth]{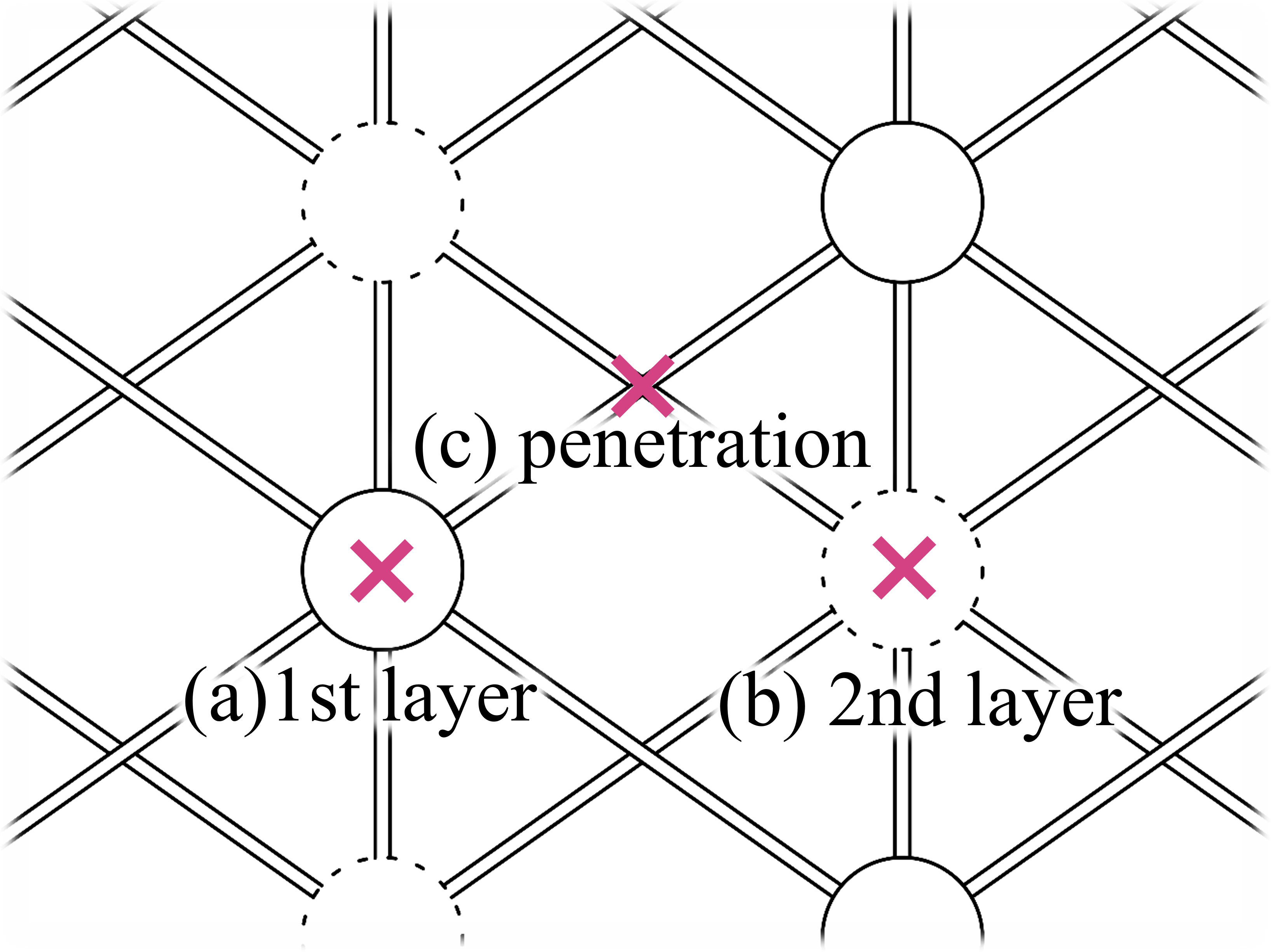}
    \caption[Three kinds of incident conditions. \added{Solid and dashed circles represent atoms of the first layer and the second layer, respectively.}
             (a) an incident from straight above the tungsten atom in the first layer, (b) an incident from straight above the tungsten atom in the second layer, and (c) an incident from the center of the two nearest atoms in the first layer. 
             In all cases, the incident angle is perpendicular to the (110) surface.]{}
    \label{fig:figure1}
\end{figure}

\begin{figure}[ht]
    \centering
    \includegraphics[width=0.7\textwidth]{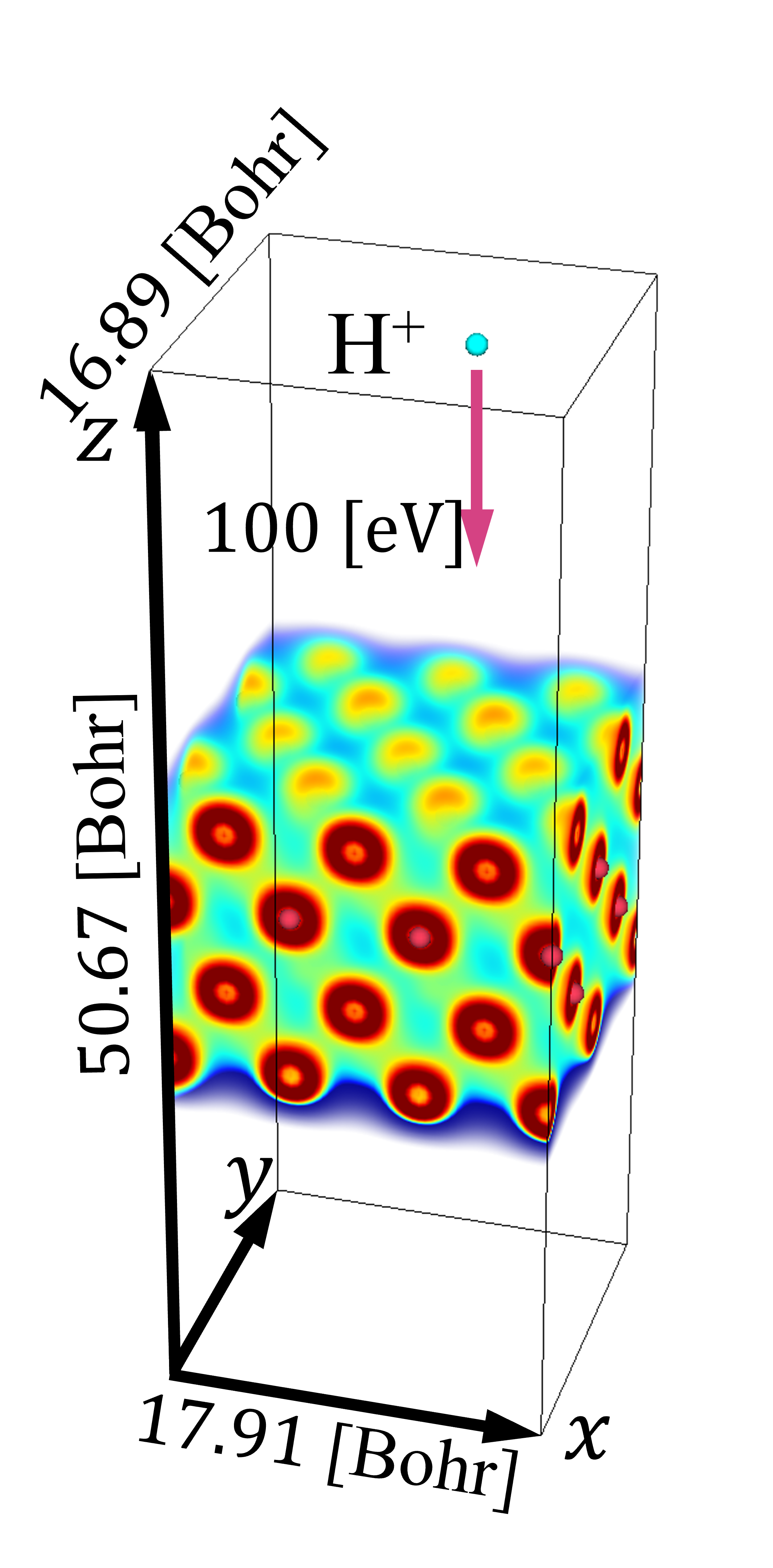}
    \caption[The simulation system. The blue and red spheres indicate the hydrogen ion and tungsten atoms, respectively.]{}
    \label{fig:figure2}
\end{figure}

The incident target is the (110) surface of the tungsten material, which is composed of $48$ atoms and $576$ electrons. 
The tungsten material has a body-centered cubic (BCC) structure, and the lattice constant is set to $3.16$ Å. 
The hydrogen ion is injected onto the surface perpendicularly. 
The incident energy is $100$ eV. 
We simulate three cases for incident positions: (a) straight above a tungsten atom in the first layer, (b) straight above a tungsten atom in the second layer, and (c) in the gap between tungsten atoms, as shown in Figure \ref{fig:figure1}. 
For cases (a) and (b), the incident ion collides with the target atom under a condition that an impact parameter is 0. 
For case (c), the incident ion penetrates due to channeling. 
The time step is $2.41 \times 10^{\added{-}19}$ sec. 
The size of the simulation box is $17.891\times16.89\times50.67$ Bohr$^3$ (see Fig. \ref{fig:figure2}), and the spatial grid is $60\times60\times180$. 
The exchange-correlation term is the local spin density approximation \cite{Perdew1981}. 
The pseudo\added{-}potential is the MBK potential \cite{Morrison1993} which is a norm-conserving type pseudo-potential generated by ADPACK \added{for OpenMX }\cite{Ozaki2003,Ozaki2004}.

The initial state of the system was generated in two steps. 
First, we calculated the relaxed structure of the tungsten surface with a vacuum region of $23.88$ Bohr using DFT. 
Next, a hydrogen ion was added to the position of the case (a), (b), or (c) in the simulation box. 
The z-coordinate of the hydrogen ion was $50.0$ Bohr for all cases, corresponding to a distance of $20.7$ Bohr from the tungsten surface. 
The initial state of the system was generated through this procedure. 
All of these simulations were performed by using the QUMASUN code \cite{Ito2024}.

\section{Analysis method for detection probabilities of electrons}

\begin{figure}[ht]
    \centering
    \includegraphics[width=0.8\textwidth]{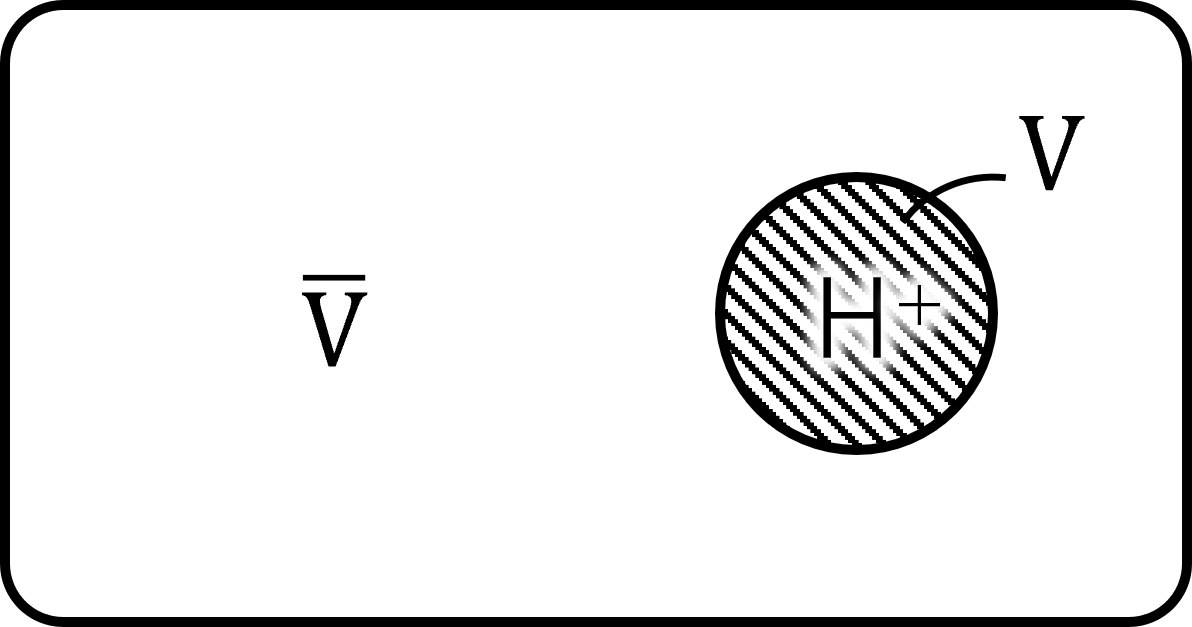}
    \caption[The hatched region represents the small region $\text{V}$ containing hydrogen, which is used in the evaluation of detection probability. The rest of the region is $\bar{\text{V}}$.]{}
    \label{fig:figure3}
\end{figure}

The aim of this chapter is to estimate the neutralization probability of the incident hydrogen after the collision with the tungsten surface from the electron wave function. 
Consider a small region ${ V}$, containing a hydrogen nucleus and let the rest of the region be $\bar{{ V}}$ (see Fig. \ref{fig:figure3}). 
If no electrons are detected inside ${ V}$, it can be considered that the hydrogen is a positive ion. 
If one electron is detected inside ${ V}$, the hydrogen is a neutral atom. 
In addition, if two electrons are detected inside ${ V}$, the hydrogen is a negative ion. 
In the quantum many-body system, the probability of detecting $m$ electrons in ${ V}$ is defined as
\begin{align}
P(m) = & \binom{n}{m}
\prod_{i=1}^{m}\int_{ V}d{r_i}
\prod_{j=m+1}^{n}\int_{\bar{V}} d{r_j} p({r_1}, {r_2}, \cdots, {r_n}).\label{eq:P_m}
\end{align}
In the case of hydrogen, $P(0)$, $P(1)$, and $P(2)$ correspond to the probabilities of detecting a positive ion, neutral atom, and negative ion, respectively. The probability density is given by 
\begin{equation}
p(r_1, r_2, \cdots, r_n) = \sum_{\sigma_1}\sum_{\sigma_2}\cdots\sum_{\sigma_n}\Psi^*(x_1, x_2, \cdots, x_n)\Psi(x_1, x_2, \cdots, x_n), \label{eq:p}
\end{equation}
where $x_i$ is the variable of a pair of the spatial coordinates $r_i$ and spin variables $\sigma_i (i=1,2,\cdots,n)$. The all-electron wave function is approximated by the Slater determinant
\begin{align}
\Psi({x_1}, {x_2},\cdots, {x_n}) = \frac{1}{\sqrt{n!}}
\begin{vmatrix}
\psi_1({x_1}) & \cdots & \psi_1({x_n}) \\
\vdots        & \ddots & \vdots      \\
\psi_n({x_1}) & \cdots & \psi_n({x_n}) \\
\end{vmatrix}\label{eq:Slater_det}
\end{align}
where $\psi_i(x)$ represents KS orbital. 
Here, the KS orbitals can be simulated by using TDDFT. 
In the present simulation, we assume that the former $a$ KS orbits are up-spin states and the latter $n-a$ KS orbits are down-spin states.

Next, the integration for $\bar{{ V}}$ can be divided into full-space and ${ V}$ components
\begin{equation}
\prod_{j=m+1}^{n}\int_{\bar{ V}}dr_j = \prod_{j=m+1}^{n}\left(\int dr_j - \int_{ V}dr_j\right)\label{eq:div_Vbar}
\end{equation}
And, we introduce
\begin{align}
\eta(s) &=
\sum_{1 \leq k_1 < k_2 < \cdots < k_s \leq n}
\sum_{\tau}{\rm sgn}\left(\tau\right)
\prod_{i = 1}^{s} \braket{\psi_{k_i}|\psi_{\tau(k_i)}}_{ V}\label{eq:eta_s}\\
\braket{\psi_i|\psi_{j}}_{ V} &= \sum_{\sigma}\int_{ V}dr \psi^*_i(x)\psi_{j}(x)\label{eq:q_ij}
\end{align}
where $\tau$ is a permutation of $(k_1, k_2, \cdots , k_s )$ to $(\tau(k_1),\tau(k_2), \cdots, \tau(k_s) )$ and $\sum_{\tau}$is the sum over all permutations. 
Consequently, using eq.(\ref{eq:div_Vbar}) and the binomial theorem, eq.(\ref{eq:P_m}) can be rewritten as
\begin{align}
P(m) 
&= \sum_{s = m}^n \frac{(-1)^{s-m}s!}{m!(s-m)!}\eta(s)\nonumber\\
&= \eta(m) + (m+1)\eta(m+1) + \frac{(m+2)(m+1)}{2}\eta(m+2) + O({\braket{\psi_i|\psi_j}_{ V}}^{m+3})\label{eq:P_m_approx}
\end{align}
Because $\eta(s)$ is the $s$-th order term of $\braket{\psi_i|\psi_j}_{ V}$. 
In the present work, $m+2$ order approximations are used.
The range of $\braket{\psi_i|\psi_j}_{ V}$ is $0 \leq \braket{\psi_i|\psi_j}_{ V} \leq 1$. 
If $\braket{\psi_i|\psi_j}_{ V}$ is sufficiently small, thus the $m+2$ order approximation provides a reasonable estimation.
In this study, the maximum value of $\braket{\psi_i|\psi_j}_{ V}$ is approximately $0.1$.
We also extend the above method to consider the probability $P(\mu,\nu)$ of detecting $\mu$ up-spin electrons and $\nu$ down-spin electrons in ${ V}$. 
$P(\mu,\nu)$ is expressed as the product of $P_\uparrow(\mu)$, the probability of detecting $\mu$ up-spin electrons in ${ V}$, and $P_\downarrow(\nu)$, the probability of detecting $\nu$ down-spin electrons
\begin{align}
P(\mu, \nu) =P_\uparrow(\mu)P_\downarrow(\nu),\label{eq:P_mn}
\end{align}
where 
\begin{align}
P_\uparrow(\mu) &= \sum_{s = \mu}^{a}\frac{(-1)^{s-\mu}s!}{\mu!(s-\mu)!}\eta_{\uparrow}(s),\label{eq:P_up_m}
\\
P_\downarrow(\nu) &= \sum_{s = \nu}^{b}\frac{(-1)^{s-\nu}s!}{\nu!(s-\nu)!}\eta_{\downarrow}(s),\label{eq:P_dw_n}
\\
\eta_{\uparrow}(s) &=
\sum_{1 \leq k_1 < k_2 < \cdots < k_s \leq a}
\sum_{\tau}{\rm sgn}\left(\tau\right)
\prod_{i = 1}^{s} \braket{\psi_{k_i}|\psi_{\tau(k_i)}}_{ V},\label{eq:eta_up_s}
\\
\eta_{\downarrow}(s) &=
\sum_{a+1 \leq k_1 < k_2 < \cdots < k_s \leq n}
\sum_{\tau}{\rm sgn}\left(\tau\right)
\prod_{i = 1}^{s} \braket{\psi_{k_i}|\psi_{\tau(k_i)}}_{ V}.\label{eq:eta_dw_s}
\end{align}
Here, $\sum_{1 \leq k_1 < \cdots < k_s \leq a}$ and $\sum_{a+1 \leq k_1 < \cdots < k_s \leq n}$represent a summation over all sets of $s$ integers $(k_1, k_2, \cdots , k_s)$ such that ${1 \leq k_1 < \cdots < k_s \leq a}$ and ${a+1 \leq k_1 < \cdots < k_s \leq n}$, respectively.

\section{Results and Discussion}

\begin{figure}[ht]
    \centering
    \includegraphics[width=0.8\textwidth]{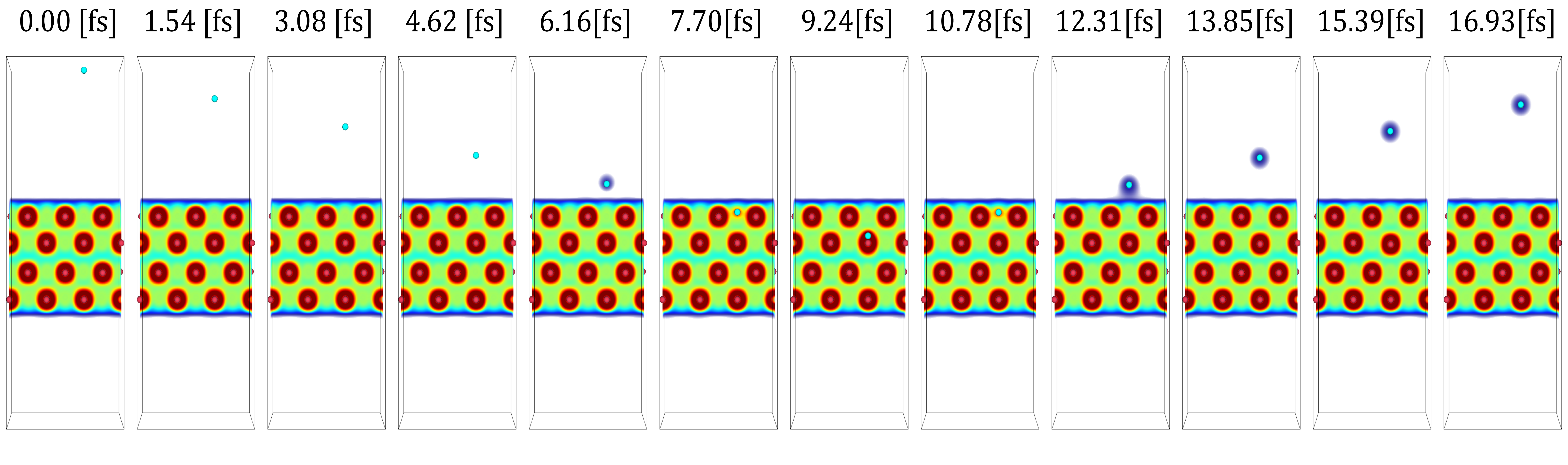}
    \caption[Snapshots of the TDDFT and MD combined simulation for the incidence of a hydrogen ion onto a tungsten surface under case(b).
             The blue and red solid spheres indicate the hydrogen and tungsten nuclei, respectively. 
             The colored area around the nucleus is the electronic density. 
             The figure is the cross-section of the simulation box parallel to the x-z plane whose y coordinate is the same as that of the hydrogen ion.]{}
    \label{fig:figure4}
\end{figure}

Figure \ref{fig:figure4} shows snapshots of the simulation for the hydrogen ion (H$^+$) incidence of the case (b). 
In the initial state, the hydrogen ion was positioned far from the tungsten surface, and there was no electronic density around the hydrogen ion. 
At $6.16$ fs, it iwas confirmed that the electronic density transferred from the side of the tungsten surface to the region near the hydrogen nucleus. 
At this time, the distance between the hydrogen and the tungsten surface was $4.59$ Bohr, which was longer than the distance to generate chemical bonding. 
After the hydrogen nucleus collided with the nearest tungsten atom in the second layer, it was reflected and escaped from the surface. 
Even when the reflected hydrogen nucleus moved far enough away from the surface, there was still an electronic density around the hydrogen. 
In cases (a) and (c) also, electronic density was transferred from the tungsten surface to the hydrogen, similar to case (b). 
In case (a), the hydrogen ion then collided with a tungsten atom in the first layer and was subsequently reflected away from the surface. 
However, in the case (c), the hydrogen penetrated the tungsten surface, which was the expected behavior from this incident condition, which was to be injected into the gap between the tungsten atoms.

\begin{figure}[ht]
    \centering
    \includegraphics[width=0.8\textwidth]{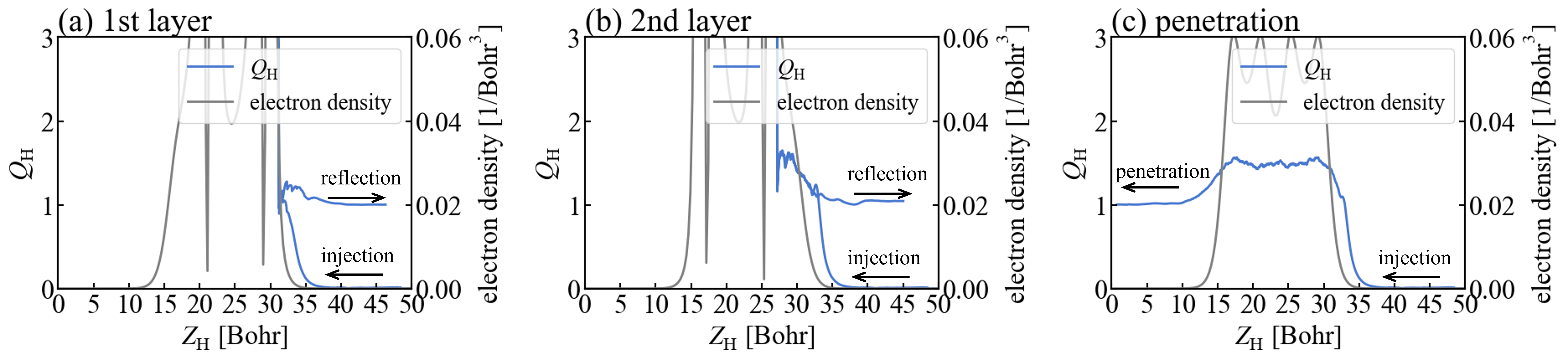}
    \caption[The blue line indicates the Bader charge $Q_{\rm H}$ around the hydrogen atom as a function of the z-coordinate of the hydrogen nucleus. 
             The black line is the initial electronic density at the positions corresponding to the hydrogen coordinates. 
             This initial electronic density is shown, even though the density changes during the simulation. 
             Figures (a), (b), and (c) correspond to the incident conditions.]{}
    \label{fig:figure5}
\end{figure}

To analyze the transition of the electronic density, the Bader charge \cite{Bader1990} which is an electronics charge around the nuclei, is estimated using the method proposed by Henkelman et, al, \cite{Henkelman2006}.
The Bader charge of the hydrogen nucleus, $Q_{\rm H}$, is shown in Fig. \ref{fig:figure5}.
\added{For all the cases of (a), (b) and (c), the position $z_{\rm H}$ decreases from 50 Bohr with the evolution of time.}
\added{When $z_{\rm H}$ reaches about 35 Bohr, the charge $Q_{\rm H}$ increases from 0 to 1 or greater.}
Note that $Q_{\rm H}$ becomes much larger within the range $30.0 < z_{\rm H} < 31.2$ in case (a) and within the range for $26.5 < z_{\rm H} < 27.2$ in case (b). 
In these regions, the distance between hydrogen and the nearest tungsten atom is very small, and the Bader volume around the hydrogen is not correctly estimated. 
After being reflected or penetrating, $Q_{\rm H}$ is approximately $1$ for all incident conditions.

\begin{figure}[ht]
    \centering
    \includegraphics[width=0.8\textwidth]{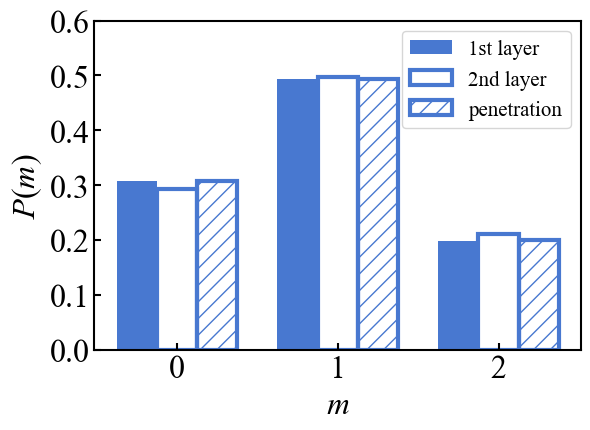}
    \caption[The detection probability of each state of hydrogen in the final state. 
             The filled, open, and hatched bars show the detection probabilities for incident conditions (a), (b), and (c), respectively.]{}
    \label{fig:figure6}
\end{figure}

Here, even though the Bader charge is close to $1$, it does not simply mean that the hydrogen has been neutralized. 
This is a problem unique to quantum many-body systems and we need to discuss it using the wave function, not the electron density.
We consider the final state of the hydrogen at the end of the simulation to be a superposition of three states: a positive ion, a neutral atom, and a negative ion. 
We can estimate the detection probability of electrons in the small region ${ V}$ using Eq. (\ref{eq:P_m_approx}). The detection probability of electrons can be interpreted as the probability of detecting a particle in each state. 
\deleted{For example, in hydrogen}\added{In the present work}, the probabilities of detecting a positive ion, a neutral atom, and a negative ion can be estimated from $P(0)$, $P(1)$ and $P(2)$, respectively. 
To estimate the probability of detecting a particle in each state, the small region ${ V}$ is defined as a simple sphere with a radius of $5.67$ Bohr, centered on the hydrogen nucleus. 
The tungsten and hydrogen are sufficiently far apart that defining the region ${ V}$ around the hydrogen as a simple sphere provides a good estimate. 
In fact, the distances between the hydrogen and the tungsten for the cases of (a), (b) and (c) are $17.17$, $15.89$, and $16.57$ Bohr, respectively, at the end of the simulations. 
As a result, Figure \ref{fig:figure6} shows that the probability of detecting a positive ion is approximately $30$ percent, a neutral atom is approximately $50$ percent, and a negative ion is approximately $20$ percent. 
These probabilities hardly depend on the incident positions. It is interesting that even though the Bader charge around the hydrogen is close to $1$, the probability of detecting a neutral hydrogen atom is at most $50$ percent.

Electrons must obey the Pauli exclusion principle. 
By estimating the probability of detecting two electrons in the same spin state, we can evaluate the contribution of the Pauli exclusion principle. 
Table \ref{table:table1} shows the detection probability of $\mu$ up-spin and $\nu$ down-spin electrons in V, evaluated using eq.(\ref{eq:P_mn}). 
Summing $P(\mu, \nu)$ for $\mu + \nu = m$ gives the detection probability of m electrons in the ${ V}$: $P(m) = \sum_{\mu = 0}^m P(\mu, m - \mu) $ . 
\added{The fact that $P(1, 0)$ and $P(0, 1)$ are equal is caused by the spin symmetry of the system.}
\added{$P(1,1)$ implies the negative ion is composed of one up-spin electron and one down-spin electron, which corresponds to spin singlet state.}
\added{Furthermore, as long as the Pauli exclusion principle is satisfied in this simulation, the probabilities $P(2,0)$ and $P(0,2)$ are the probabilities of a spin triplet state.}
\added{Actually, the fact that}\deleted{The fact that} $P(0,2)$ and $P(2,0)$ are approximately zero\deleted{ indicates that there is a very small probability of detecting two electrons with the same spin state within ${ V}$}. 
\deleted{$P(2,0)$ and $P(0,2)$ are the sum of all combinations of the probability of two electrons in the same spin state occupying all two electronic orbitals. 
In other words, $P(0, 2)$ and $P(2, 0)$ include the probability of detecting two electrons in the same spatial orbit and spin that. 
Therefore, if $P(0, 2) \approx P(2, 0) \approx 0$, it indicates that the probability of detecting two electrons with the same spin in the same spatial orbital is also zero. 
This result satisfies the the Pauli exclusion principle and supports the validity of the analysis method. 
Moreover, it indicates that the probability is approximately 0 for two electrons to occupy both the 1s and 2s orbitals in the same spin state as one of the spin-triplet states.}
\deleted{The fact that $P(1, 0)$ and $P(0, 1)$ are equal is caused by the spin symmetry of the system.}

Previous models \cite{Horiguchi1978,Hagstrum1961,Trubnikov1967,Kitagawa1976} have not considered the formation of negative ions on a surface. 
However, it is \added{experimentally} known that under certain conditions the probability of detecting hydrogen reflected as a negative ion is significantly higher than that of detecting a positive ion\cite{Verbeek1980}. 
\added{In addition, these models don't take into account electron spin.}
\added{In contrast}\deleted{On the other hand}, we performed TDDFT simulations \added{with spin and demonstrated the generation of not only a neutral atom but also a negative ion.}\deleted{that included all states of hydrogen after a collision, not only neutral or positive ions.} 
As a result, the probability of detecting negative hydrogen was \added{not small.}\deleted{significantly high, and this concern needed to be taken into account.}
\added{This fact expects that the TDDFT simulation can contribute to the future clarification of the mechanism of negative ion generations.}

\section{Conclusion}

For the neutralization process of the hydrogen ion on tungsten surfaces, which is a historical problem in the research field on PWI, we performed TDDFT simulation to treat the quantum transition process of electron wave functions. 
As a result, the state of the hydrogen after the collision with the surface was obtained as the superposition of the following three states: the reflected hydrogen became a positive ion, a neutral atom, or a negative ion. 
We also proposed a method to evaluate the detection probabilities of those three states, which corresponds to the probability of detecting zero, one or two electrons around the hydrogen nucleus. 
As a result, the Bader charge of the hydrogen after reflection was approximately 1. 
However, the probability of detecting the hydrogen as a neutral atom was approximately 50 percent. 
In addition, it was also possible to estimate the probability of detecting the hydrogen as a negative ion\added{ by considering electron spin}.

\section*{Acknowledgments}
The present work was supported by JSPS KAKENHI Grant Numbers JP23K17679, JP24H02251, JP24K00617 and was supported by the SOKENDAI Student Dispatch Program(2024). 
The numerical simulations of TDDFT were performed on Plasma Simulator (NEC SX-Aurora TSUBASA) at NIFS with the support and under the auspices of the NIFS Collaboration Research program (NIFS24KISM002).

\clearpage
\listoffigures 
\clearpage
\section*{Figures}

\clearpage

\clearpage

\clearpage

\clearpage

\clearpage

\clearpage
\listoftables 
\clearpage
\section*{Tables}
\begin{table}[ht]
  \centering
  \caption{Detection probability of electron considering spin polarization. The table (a), (b) and (c) correspond to the incident conditions.}
  \begin{subtable}[t]{0.9\textwidth}
      \centering
      \caption{1st layer}
      \begin{tabular}{lll}
          \toprule
          $P(0) = 0.307$ &	$P(1) = 0.494$	& $P(2) = 0.199$ \\
          \midrule
          $P(0, 0) = 0.307$ &	$P(0, 1) = 0.247$ &  $P(1, 1) = 0.198$\\
	                          & $P(1, 0) = 0.247$ &	 $P(0, 2) = 2.59\times10^{-5}$\\
		                        &                 &  $P(2, 0) = 2.59\times10^{-5}$\\
          \bottomrule
      \end{tabular}
  \end{subtable}
  \vspace{0.5cm}
  \begin{subtable}[t]{0.9\textwidth}
      \centering
      \caption{2nd layer}
      \begin{tabular}{lll}
          \toprule
          $P(0) = 0.293$	& $P(1) = 0.497$	& $P(2) = 0.211$\\
          \midrule
          $P(0, 0) = 0.293$ & $P(0, 1) = 0.248$ & $P(1, 1) = 0.211$\\
	                        & $P(1, 0) = 0.248$ & $P(0, 2) = \added{3.20}\times10^{-5}$\\
		                    &                   & $P(2, 0) = \added{3.20}\times10^{-5}$\\
          \bottomrule
      \end{tabular}
  \end{subtable}
  \vspace{0.5cm}
  \begin{subtable}[t]{0.9\textwidth}
      \centering
      \caption{penetration}
      \begin{tabular}{lll}
          \toprule
          $P(0) = 0.306$	& $P(1) = 0.494$	& $P(2) = 0.199$\\
          \midrule
          $P(0, 0) = 0.306$	& $P(0, 1) = 0.247$	& $P(1, 1) = 0.199$\\
                            & $P(1, 0) = 0.247$	& $P(0, 2) = 2.39\times10^{-5}$\\
                            &                   & $P(2, 0) = 2.39\times10^{-5}$\\
          \bottomrule
      \end{tabular}
  \end{subtable}
  \label{table:table1}
\end{table}

\end{document}